\documentclass[aps,twocolumn,nature,epsfig]{revtex4}

\usepackage{epsf}
\usepackage{amsfonts}
\usepackage{amssymb}
\usepackage{graphicx}
\usepackage{color}
\usepackage{amsmath}
\usepackage[bookmarksnumbered, bookmarks, breaklinks, linktocpage]{hyperref}

\newcommand{\Sys}{\ensuremath{\mathcal{S}}}
\newcommand{\Env}{\ensuremath{\mathcal{E}}}

\def\FCW{0.98\columnwidth}

\newcommand{\bra}[1]    {\langle #1|}
\newcommand{\ket}[1]    {| #1 \rangle}
\newcommand{\bk}[2]     {\langle #1 | #2 \rangle}
\newcommand{\kb}[2]     {| #1 \rangle \! \langle #2 |}
\newcommand{\cH}        {{\mathcal H}}
\newcommand{\cS}        {{\mathcal S}}

\newcommand{\cE}        {{\mathcal E}}
\newcommand{\cN}        {{\mathcal N}}

\newcommand\cF{{\mathcal F}}
\newcommand\hocom[1]{}

\newcommand{\ba}{\begin{eqnarray}}
\newcommand{\ea}{\end{eqnarray}}
\newcommand{\bmath}{\begin{mathletters}}
\newcommand{\emath}{\end{mathletters}}
\newcommand{\ban}{\begin{eqnarray*}}
\newcommand{\ean}{\end{eqnarray*}}

\begin{document}

\title{Quantum Darwinism}


\author{Wojciech Hubert Zurek}

\address{Theory Division, MS B213, LANL
    Los Alamos, NM, 87545, U.S.A.}


\begin{abstract}
Quantum Darwinism describes the proliferation, in the environment, of multiple records of selected states of a quantum system. It explains how the fragility of a state of a single quantum system can lead to the classical robustness of states of their correlated multitude; shows how effective `wave-packet collapse' arises as a result of proliferation throughout the environment of imprints of the states of quantum system; and provides a framework for the derivation of Born's rule, which relates probability of detecting states to their amplitude. Taken together, these three advances mark considerable progress towards settling the quantum measurement problem.
\end{abstract}
\maketitle



The quantum principle of superposition implies that any combination of quantum states is also a legal state. This 
seems to be in conflict with everyday reality: States we encounter are localized. Classical objects can be either here or there, but never {\it both} here and there. Yet, the principle of superposition says that localization should be a rare exception and not a rule for quantum systems. 

Fragility of states is the second problem with quantum-classical correspondence: Upon measurement, a general preexisting quantum state is erased -- it ``collapses'' into an eigenstate of the measured observable. How is it then possible that objects we deal with can be safely observed, even though their basic building blocks are quantum? 

To bypass these obstacles Bohr~\cite{11} followed Alexander the Great's example: Rather than try disentangling the Gordian Knot at the beginning of his conquest, he cut it. The 
cut 
separates the quantum from the classical. Bohr's Universe consists of two realms, each governed by its own laws. Fragile 
superpositions were banished from the classical realm deemed more fundamental and indispensable to interpret or even practice quantum theory. Thus, instead of trying to understand  Universe (including ``the classical'') in quantum terms 
one  ``quantized'' 
this and that, always 
starting from the classical base.

This was a brilliant tactical move: Physicists could conquer the quantum realm without getting distracted by
interpretational worries. In those days only gedankenexperiments like the famous Schr\"odinger cat~\cite{47} were truly disturbing: Real experiments dealt with electrons, photons, atoms, or other microscopic systems. Bohr's rule of thumb -- that the macroscopic is classical -- was enough. Moreover, many (including Einstein) believed that quantum physics is just a step on a way to a deeper theory that will solve or bypass interpretational conundrums. 

That did not happen. Instead, old gedankenexperiments were carried out. They confirmed validity of quantum laws on scales that have, of recent, begun to infringe on ``the macroscopic''. Quantum theory is here to stay. It is also increasingly clear that its weirdest predictions -- superpositions and entanglement -- are experimental facts, in principle relevant also for macroscopic objects. Therefore, questions about the origin of ``the classical'', with its restriction to localized states that are robust, unperturbed by measurements, can no longer be dismissed.

\section{Decoherence and einselection}

Decoherence 
turns one of the two problems we noted above -- fragility of quantum states -- into a solution of the other. Environment-induced decoherence recognizes that 
if a measurement can put a state at risk and re-prepare it, so can accidental information transfers that happen whenever a system interacts with its environment. 

Decoherence is by now well understood~\cite{36,75,52}: Fragility of states makes quantum systems very difficult to isolate. Transfer of information (which has no effect on classical states) has dramatic consequences in the quantum realm. 
So, while fundamental problems of classical physics were always solved in isolation (it sufficed to prevent energy loss) this is not so in quantum physics (leaks of information are much harder to plug). 

When a quantum system gives up information, its own state becomes consistent with the information that was disseminated. ``Collapse'' in measurements is an extreme example, but any interaction that leads to a correlation can contribute to such re-preparation: Interactions that depend on a certain observable correlate it with the environment, so its eigenstates are singled out, and phase relations between such {\it pointer states} are lost~\cite{69}.

Negative selection due to decoherence is the essence of {\it e}nvironment-{\it in}duced super{\it selection}, or {\it einselection}~\cite{70}: Under scrutiny of the environment, only pointer states remain unchanged. Other states decohere into mixtures of stable pointer states
that can persist, and, in this sense, exist: They are einselected.

These ideas can be made precise. The basic tool is the reduced density matrix $\rho_{\cS}$. It represents the state of the system that obtains from the composite state ${\Psi_{\cS\cE}}$ of $\cS$ and $\cE$ by tracing out the environment $\cE$:
$$ \rho_\cS =
Tr_{\cE} \kb {\Psi_{\cS\cE}}  {\Psi_{\cS\cE}} \ . \eqno(1)$$
Evolution of $\rho_\cS$ reveals preferred states: It is most predictable when the system starts in a pointer state. To quantify this one can use (von Neumann) entropy,
$ H_{\cS}=H(\rho_\cS) = - Tr \rho_\cS \lg \rho_\cS$,
as a function of time. Pointer states 
result in smallest entropy increase. By contrast, their superpositions produce entropy rapidly, at decoherence rates, especially when $\cS$ is macroscopic. 

When pure states of the system are sorted by predictability, according to entropy of the evolved $\rho_\cS$, pointer states are at the top. This criterion -- the {\it predictability sieve}~\cite{75,45,80} -- yields a short list of candidates for effectively classical states: A cat can persist in one of the two obvious stable states, but their superposition would deteriorate into a mixture of $\ket {\tt dead}$ and $\ket {\tt alive}$ 
when initiated in a way envisaged by Schr\"odinger~\cite{47}. 

The special role of position is traced to the nature of the $\cS \cE$ interactions: They tend to depend on distance. Hence, information about position is most readily passed on to the environment. This is why localized states survive while nonlocal superpositions decay into their mixtures. For example, in a weakly damped harmonic oscillator the minimum uncertainty wavepackets -- familiar coherent states, best quantum approximation of classical points in phase space -- are einselected~\cite{80,30,55}. 

\section{Environment as a witness}

Monitoring by the environment means that information about $\cS$ is deposited in $\cE$. What role does it play, and what is its fate? Decoherence theory ignores it. Environment is ``traced out''. Information it contains is treated as inaccessible and irrelevant: $\cE$ is a ``rug to sweep under'' the data that might endanger classicality.

Quantum Darwinism recognizes that ``tracing out'' is not what we do: Observers eavesdrop on the environment. Vast majority of our data comes from fragments of $\cE$.  Environment is a {\it witness} to the state of the system. 

For example, this very moment {\it you} intercept a fraction of the photon environment emitted by a screen or scattered by a page. We never access all of $\cE$. Tiny fractions suffice to reveal the state of various ``systems of interest''.  

This insight captures the essence of Quantum Darwinism: Only states that produce multiple informational offspring -- multiple imprints on the environment -- can be found out from small fragments of $\cE$. The origin of the emergent classicality is then not just survival of the fittest states (the idea already captured by einselection), but their ability to ``procreate'', to deposit multiple records -- copies of themselves -- throughout $\cE$.

\begin{figure}[tb]
\begin{tabular}{l}
\vspace{-0.15in} \includegraphics[width=\FCW]{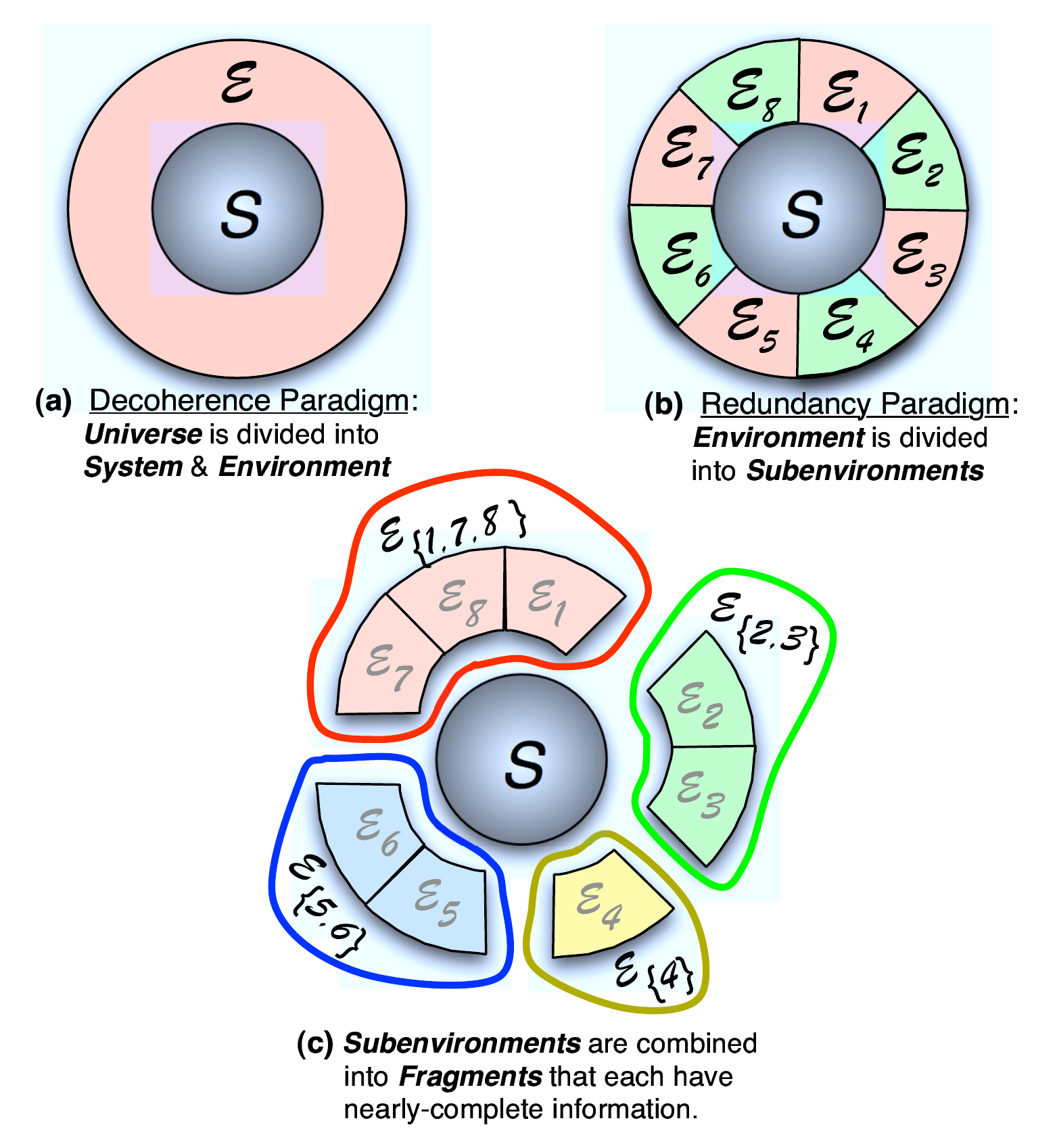}\\
\end{tabular}
\caption{\emph{Quantum Darwinism and the structure of the environment}.  Decoherence theory distinguishes between a system ($\Sys$) and its environment ($\Env$) as in \textbf{(a)}, but makes no further recognition of the structure of $\cE$; it could as well be monolithic.  In Quantum Darwinism the focus is on {\it redundancy}. We recognize the subdivision of $\Env$ into subsystems, as in \textbf{(b)}. The only 
requirement for a subsystem is that it should be individually accessible to measurements; observables of different subsystems commute. To obtain information about $\cS$ from $\cE$ one can then measure \emph{fragments} $\cF$ of the environment -- non-overlapping collections of subsystems of $\cE$, \textbf{(c)}.  
ically, there are many copies of the information about $\cS$ in $\cE$ -- 
``progeny'' of the ``fittest observable'' that survived monitoring by $\cE$ proliferates throughout $\cE$. This proliferation of the multiple informational offspring defines Quantum Darwinism. 
The environment becomes a witness with redundant copies of information about the preferred
observable. This leads to the objective existence of pointer states: Many can find out the state of the system independently, without prior information, and they can do it indirectly, without perturbing $\cS$.}
\label{EnvSubdivision}
\end{figure}

Proliferation of records allows information about $\cS$ to be extracted from many fragments of $\cE$ (in the example above, photon $\cE$). Thus, $\cE$ acquires {\it redundant} records of $\cS$.  Now, many observers can find out the state of $\cS$ independently, and without perturbing it. This is how preferred states of $\cS$ become objective. Objective existence -- hallmark of classicality -- emerges from the quantum substrate as a consequence of redundancy. 

Decoherence theory was focused on the system. Its aim was to determine what states survive information leaks to $\cE$. Now we ask: What information about the system can be found out from fragments of $\cE$?  
This change of focus calls for a more realistic model of the environment (Fig.~1): Instead of a monolithic $\cE$ we recognize that environments consist of subsystems that comprise {\it fragments}  independently accessible to observers.

The reduced density matrix $\rho_\cS$ representing the state of the system was the basic tool of decoherence. To study Quantum Darwinism we focus on correlations between fragments of the environment and the system. The relevant reduced density matrix $\rho_{\cS\cF}$ is given by:
$$ \rho_{\cS\cF} 
= Tr_{\cE/\cF} \kb {\Psi_{\cS\cE}}  {\Psi_{\cS\cE}} \ . \eqno(2)$$
Above, trace is over ``$\cE$ less $\cF$'', or ${\cE/\cF}$  -- all of $\cE$ except for the fragment $\cF$.
How much $\cF$ knows about $\cS$ can be quantified using mutual information:
$$I(\cS : \cF) = H_{\cS}+ H_{\cF} - H_{\cS, \cF} \ , \eqno(3) $$
defined as the difference between entropies of two systems (here $\cS$ and $\cF$) treated separately and jointly. For example, the mutual information between an original and a perfect copy (of, say, a book) is equal to the entropy of the original, as either contains the same text. So, every bit of information in the first copy reveals a bit of information in the original. However, 
having extra copies does not increase the information about the original. Yet, it determines how many can independently access this information. The number of copies defines redundancy. 

\begin{figure}[tb]
\begin{tabular}{l}
\vspace{-0.01in} \includegraphics[width=\FCW]{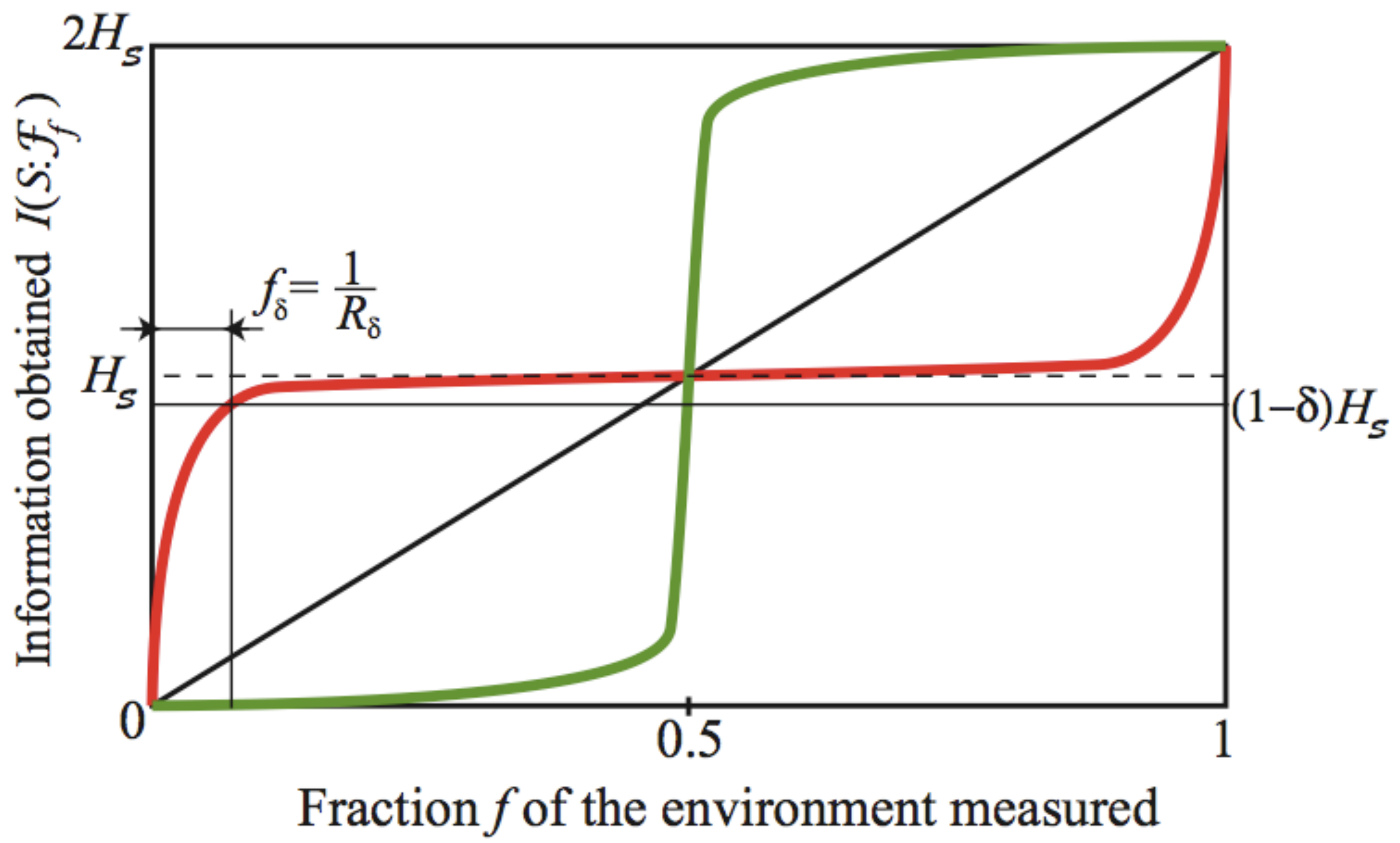}\\
\end{tabular}
\caption{\emph{Information about $\cS$ stored in $\cE$ and its redundancy}. Mutual information is monotonic in $f$. When global state of $\cS\cE$ is pure, $I(\cS:\cF_f)$ in a typical fraction $f$ 
of the environment is antisymmetric around $f=0.5$~\cite{8}. For pure 
states picked out at random from the combined Hilbert space $\cH_{\cS\cE}$, there is little mutual information between $\cS$ and a typical $\cF$ smaller than half of $\cE$. However, once a 
threshold $f=\frac 1 2$ is attained, nearly all information is in principle at hand. Thus, such random 
states (green line) exhibit no redundancy. By contrast, states of $\cS\cE$ created by decoherence (where the environment monitors preferred observable of $\cS$) contain almost all (all but $\delta$) of the information about $\cS$ in small fractions $f_\delta$ of $\cE$. The corresponding $I(\cS:\cF_f)$ (red line) quickly rises to $H_\cS$ (entropy of $\cS$ due to decoherence), which is all 
of the information about $\cS$ available from either $\cE$ or $\cS$. (More, up to $2H_\cS$, can be obtained only through global measurements on $\cS$ and nearly all $\cE$). $H_\cS$ is therefore the 
\emph{classically accessible information}. As $(1-\delta)H_\cS$ of information is contained in
$f_\delta=1/R_\delta$ of $\cE$, there are $R_\delta$ such fragments in $\cE$: $R_\delta$ 
is the {\it redundancy} of the information about $\cS$. Large redundancy implies objectivity: The state 
of the system can be found out indirectly and independently by many observers, who will agree about 
their conclusions. 
Thus, \emph{Quantum Darwinism accounts for the emergence of objective existence}.}
\label{SnapPIP}
\end{figure}

Similar ideas apply to the quantum case. Initially, every bit of information gained from a fraction $f\ll1$ of $\cE$ that was pure before it monitored (and decohered) the system is a bit about $\cS$. The red plot in Fig.~2 starts with this steep ``bit for bit'' slope, but moderates as $I(\cS:\cF_f)$ approaches redundancy plateau at $H_{\cS}$, where additional bits only confirm what is already known.

Redundancy is the number of independent fragments of the environment that supply Òalmost allÓ classical information about  $\cS$, i.e., $(1-\delta) H_\cS$. In other words;
$$ R_\delta=1/f_\delta \ . \eqno(4)$$
$ R_\delta$ is the number of times one can acquire $(1-\delta)$ of the information about $\cS$ independently (from distinct $\cF$'s) and indirectly -- without perturbing $\cS$.

Rapid rise and gradual leveling of  $I(\cS:\cF_f)$, Fig. 2, implies redundancy. The information 
in $\cF_f$ allows one to determine the state of $\cS$  as it reaches redundancy plateau. Observables of different $\cF$'s commute -- such measurements are independent. Yet, underlying correlations mean that their outcomes imply the same state of the system, as if $\cS$ were classical: The redundancy plateau is a classical plateau. Its level $H_{\cS}$ is the classical information accessible from a small fraction of $\cE$. 

Redundancy 
allows for 
objective existence of the state of $\cS$: It can be found out indirectly, so there is no danger of 
perturbing $\cS$ with a measurement. Error correction allowed by redundancy is also important: 
Fragility of quantum states means that copies in $\cF$'s are damaged by measurements (we destroy photons!), and may be measured in a ``wrong'' basis. One cannot access records in $\cE$ without endangering their existence. But with many ($R_\delta$) copies, state of $\cS$ can be found out by $\sim R_\delta$ observers who can get their information independently, and without prior 
knowledge about $\cS$. Consensus between copies suggests objective existence of the state of $\cS$. 

The mutual information $I(\cS:\cF_f)$ computed in models of decoherence exhibits
behavior illustrated by the red plot of Fig. 2. In the family of models representing spin $\cS$ surrounded by environments of many spins~\cite{42,8,9} the same number of spins suffices to reach the plateau: Adding more spins to $\cE$ only extends length of the plateau measured in ``absolute units'' -- in the number of the environment spins. In this model (that can be viewed as a simplified model of a photon environment) redundancy is then proportional to the number of the environment subsystems that interact with the system of interest. 

Quantum Brownian motion -- harmonic oscillator surrounded by many environmental oscillators --
is the other well known model of decoherence. It is exactly solvable, and the case of an underdamped oscillator yields surprisingly simple results~\cite{10,PR}: (i) Mutual information is approximately given by $I(\cS : \cF) \approx H_{\cS} + \frac 1 2 \ln {\frac f {(1-f)}}$, 
and; (ii) Redundancy for an initially squeezed state of $\cS$ reaches $R_{\delta} \approx s^{2 \delta}$, where 
$s$, the squeeze factor, quantifies delocalization of the state. Similar equation 
should hold for more general ``Schr\"odinger cat'' states, with $s$ quantifying the separation of the two localized alternatives.
 
These results confirm intuitions that originally motivated Quantum Darwinism~\cite{75,Z2000}: Monitoring of the system by the environment can deposit multiple records of preferred states of $\cS$ in $\cE$. States of $\cS\cE$ that arise from decoherence are special~\cite{8,9}, as $I(\cS:\cF_f)$ for a typical pure state selected with Haar measure in the whole Hilbert space of $\cS\cE$ (green plot in Fig. 2) shows. In such random states small fragments reveal almost nothing about the rest of the state. Only when half of $\cE$ is found out the whole state is suddenly revealed.

States that arise from decoherence are then far from random. Roughly speaking, they have a {\it branch structure}.
This is why the rest of such a {\it branch} including the state of the system -- the ``bud'' from which this branch has originated -- can be deduced from its fragment. We shall see how such branches grow
in the next section.

Plots of $I(\cS:\cF_f)$ for pure $\cS\cE$ are antisymmetric around the point $\{H_{\cS}, f= \frac 1 2 \}$ for typical fragments of $\cE$~\cite{8}. Thus, rapid rise for small $f$ must be matched at the other end, for $f\sim 1$. This is a signature of entanglement that allows state to be known ``as the whole'', while states of subsystems are unknown. The joint state of $\cS\cE$ is then pure, so that $H_{\cS, \cF = \cE}=0$, and $I(\cS:\cF_f)$ must rise to $H_{\cS} + H_{\cE} = 2H_{\cS}$ when $f$ approaches 1. 

This is a very quantum aspect of information. In classical physics knowing a composite object implies knowing each of its subsystems. This is not so in quantum physics, where composite states are given by tensor (rather than Cartesian) products of their constituents. Thus, one can know perfectly quantum state of the whole, but know nothing about states of parts. We shall see in Section IV how this feature can be used to derive Born's rule~\cite{12} that relates probabilities with wavefunctions. 

To reveal this latent quantumness one would have to measure the right global observable on all of $\cS\cE$. For example, when mutual information, Eq.~(3), is defined using Shannon entropy with probabilities corresponding to optimal observables in $\cS$ and in $\cE$, the resulting Shannon 
$I(\cS:\cF_f)$ graph for small $f$ would look very similar to Fig.~2. However, using Shannon entropy
involves local probabilities (precluding global observables), so such Shannon $I(\cS:\cF_f)$ never exceeds $H_{\cS}$, antisymmetry is lost, and the plateau continues until the end at $f\sim 1$.

Effective unattainability of the $f\sim1$ part of the plot also shows why decoherence is so hard to undo: Correlations that reveal coherence can be usually detected only by such global measurements of whole $\cS\cE$. We intercept small fractions of $\cE$, and never have the luxury of perfect global measurements needed to undo decoherence.  Yet, because of redundancy, we get $\sim H_{\cS}$ information with ``sloppy'' measurements of $f\ll1$.

Quantum Darwinism does not require pure $\cE$.  Mixed environment is a noisy communication channel: Its initial entropy of $h$ per bit can still increase after interaction with $\cS$, reflecting mutual information buildup. However, now a bit gained from $\cE$ yields only $1-h$ of a bit about $\cS$.  So, a completely mixed $\cE$ ($h=1$) is useless (even though it can still induce decoherence!). For a partly mixed $\cE$ mutual information will increase more slowly, pure case ``bit per bit'' rate tempered to $\sim1-h$. Yet, it can still climb the same redundancy plateau at $H_{\cS}$~\cite{QZZ}.  


These conclusions apply when $\cE$ is initially mixed, but are also relevant when this channel is noisy for other reasons (e.g., imperfect measurements). In all such cases one can still reach the same redundancy plateau, although now a proportionally larger fragment of the environment 
is needed to get the same information about $\cS$. 

Suitability of the environment as a channel depends on whether it provides a direct and easy access to the records of the system. This depends on the structure and evolution of $\cE$. Photons are ideal in this respect: They interact with various systems, but, in effect, do not interact with each other. 
This is why light delivers most of our information. Moreover, photons emitted by the usual sources (e.g., sun) are far from equilibrium with our surroundings. Thus, even when decoherence is dominated by other environments (e.g., air) photons are much better in passing on information they acquire while ``monitoring the system of interest'': Air molecules scatter from one another, so that whatever record they may have gathered becomes effectively undecipherable. 

Stability of the level of the redundancy plateau at $H_{\cS}$, even for mixed $\cE$'s, is a compelling reason to think of it as ``classical''. 
The question we shall now address concerns the nature of that information -- what does the environment know about the system, and why?

\section{From Copying to Quantum Jumps}

Quantum Darwinism leads to appearance, in the environment, of multiple copies of the state of the system. However, the no-cloning theorem~\cite{24,65} 
prohibits copying of unknown quantum states. If cloning is outlawed, how can
redundancy seen in Fig. 2 be possible? 

\begin{figure*}[tb] 

\center \includegraphics[width=5.8in]{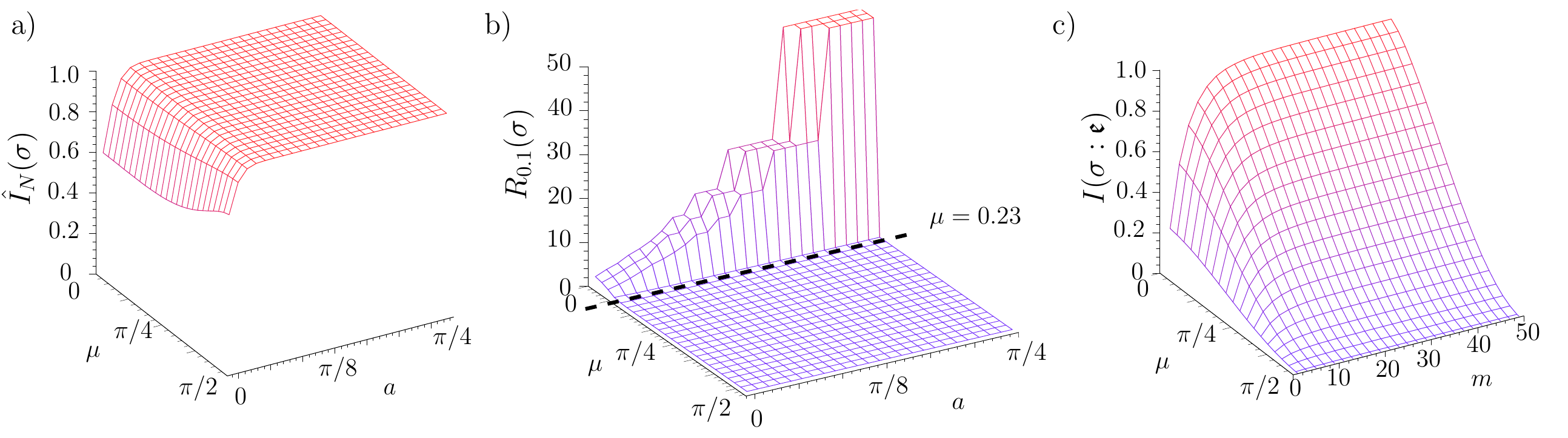}
\vspace{-0.1in}

\caption{{\it Quantum Darwinism in a simple model of decoherence}~\cite{42}. The spin-$\frac 12$ $\cS$ interacts with $N=50$ spin-$\frac 12$ subsystems of $\cE$
 with an Ising Hamiltonian ${\bf H_{\cS\cE}} = \sum_{k=1}^N g_k \sigma_z^{\cS}
  \otimes\sigma_y^{\cE_k}$. The initial state of
  $\mathcal S\otimes\cE$ is $\frac{1}{\sqrt 2}(\ket 0 + \ket 1)
  \otimes \ket 0 ^{\cE_1}\otimes\ldots\otimes\ket 0^{\cE_N}$. 
  Couplings $g_k$ are distributed randomly 
  in the interval (0,1]. All the plotted quantities are a function of 
the observable
  $\sigma(\mu) = \cos(\mu)\sigma_z + \sin(\mu)\sigma_x$, where $\mu$
  is the angle between its eigenstates and the pointer states of
  $\cS$ -- eigenstates of $\sigma_z^\cS$. {\bf a)}~Information
  acquired by the optimal measurement on the whole environment, 
  $\hat I_N(\sigma)$, as a function of the inferred observable $\sigma(\mu)$ 
  and the average interaction action $\langle g_k t \rangle = a$.
   A lot of information is accessible in the {\em whole}
$\cE$ about any observable $\sigma(\mu)$ except when 
$a$ is so small that there was no decoherence. 
  {\bf b)}~Redundancy of the information about $\cS$
  as a function of the inferred observable $\sigma(\mu)$ and the
  average action $\langle g_k t \rangle= a$. 
  $R_{\delta=0.1}(\sigma)$
  counts the number of times 90\% of
  the total information can be ``read off'' independently by measuring
  distinct fragments of $\cE$.  
  It is sharply peaked around the
  pointer observable: Redundancy is a very selective criterion -- the
number of copies of relevant information 
  is high only for the
  observables $\sigma(\mu)$ inside the theoretical bound (see
Ref.\cite{42}) indicated by the dashed line.  {\bf c)}~Information about
  $\sigma(\mu)$ extracted by 
  local random
  measurements on $m$ environmental subsystems.  
  Because of redundancy, pointer states -- and only pointer
  states -- can be found out through this far-from-optimal 
  strategy. Information about any other observable $\sigma(\mu)$ is
  restricted to 
 what can be inferred from the pointer observable~\cite{42}.}
\label{opz}
\end{figure*}
 
Quick answer is that cloning refers to (unknown) quantum {\it states}. So, copying of {\it observables} evades the theorem. Nevertheless, the tension between the prohibition on cloning and the need for copying is revealing: It leads to breaking of unitary symmetry implied by the superposition principle, 
accounts for quantum jumps, and suggests origin of the ``wavepacket collapse'', setting stage 
for the study of quantum origins of probability in Section IV.

Quantum physics is based on several ``textbook'' postulates~\cite{23}. The first two; (i) {\it States are represented by vectors in Hilbert space}, and; (ii) {\it Evolutions are unitary} -- give complete account of 
{\it mathematics} of quantum theory, but make no connection with {\it physics}. For that one 
needs to relate calculations made possible by the {\it superposition principle} of (i) and {\it unitarity} of (ii) 
to experiments.

Postulate (iii) {\it Immediate repetition of a measurement yields the same outcome} starts this task. 
This is the only uncontroversial measurement postulate (even if it is difficult to approximate in the laboratory): Such {\it repeatability} or {\it predictability} is behind the very idea of ``a state''. 

In contrast to (i)-(iii), {\it collapse postulate} (iv) {\it Outcomes correspond to eigenstates of the measured observable, and only one of them is detected in any given run of the experiment}, is inconsistent with (i) and (ii). Conflict arises for two reasons: Restriction to a preferred set of outcome states seems at odds with with the egalitarian principle of superposition, embodied in (i). This restriction prevents
one from finding out unknown quantum states, so it is responsible for their fragility.
And a single outcome per run is at odds with unitarity (and, hence, linearity) of quantum dynamics that preserves superpositions. 

The last axiom; (v) {\it Probability of an outcome is given by the square of the associated amplitude}, 
$p_k=|\psi_k|^2$, is known as {\it Born's rule}~\cite{12}.
It completes the relation between mathematics of (i) and (ii) and the experiments.

Bohr bypassed conflict of (i) and (ii) with (iv) 
by insisting that apparatus is classical, so unitarity and the principle of superposition need 
not apply to measurements. But this is an excuse, not an explanation. 
We are dealing with a quantum environment, and redundancy of previous section 
strengthened
motivation for postulate (iii) -- repeatability. Let us see 
where this demand takes us in a purely quantum setting of postulates (i), (ii), and (iii).

Suppose 
there are states 
of $\cS$ (say, $\ket u$ and $\ket v$) that produce an imprint in a subsystem of $\cE$ (which 
plays a role of an apparatus), but remain unperturbed (so they can produce more imprints). This {\it repeatability} implies:
$ \ket u \ket {e_0} \Rightarrow \ket u \ket {e_u} $,
$ \ket v \ket {e_0} \Rightarrow \ket v \ket {e_v}$
in obvious notation. In a unitary process scalar product is preserved. Thus;
$$ \bk u v = \bk u v \bk {e_u}{e_v} \ , \eqno(5)$$
where we have set $\bk {e_0} { e_0 } =1$. This simple equation can be satisfied only when; (a) either $\bk {e_u}{e_v}=1$ (which means that copying was completely unsuccessful), or; (b) $\bk u v=0$, i.e., they are orthogonal. In that case $\bk {e_u}{e_v}$ is arbitrary -- perfect record $\bk {e_u}{e_v}=0$ is also possible. 

It follows that multiple (perfect or imperfect) copies 
of $\ket u$ and $\ket v$ can be imprinted in disjoint $\cF$'s. 
As a consequence of unitarity, 
only sets of orthogonal states (that define Hermitean observables~\cite{23}) can be so copied,
explaining selection of a set of outcomes -- terminal points of quantum jumps~\cite{79}. Before, 
they had to be postulated by the first part of axiom (iv). We emphasize that this result relies on just two values of the scalar product -- 0 and 1 -- and, thus, does not appeal
to Born's rule.

This breaking of unitary symmetry (choice of preferred states in an egalitarian Hilbert space) is induced by repeatability of the information transfer.
It is a ``nonlinear demand'': As in cloning, one asks for 
``two (or more) of the same''. Its conflict with linearity of quantum theory can be resolved only by restricting states that can be copied. Such pointer states then act as ``buds'' of 
branches that grow by reproducing, in $\cE$, multiple copies of the original in $\cS$.  Interaction Hamiltonians 
do not perturb observables that commute with them. So, buds of branches coincide with the einselected pointer states.

Evidence of such symmetry breaking is seen in Fig. 3. Mutual information and redundancy shown there
are obtained using Eq. (3), but with Shannon (rather than von Neumann) entropies of specific observables of $\cS$ and $\cF$, i.e.,  
using probabilities of their eigenstates.
While
von Neumann-based $I(\cS:\cF_f)$ and $R_{\delta}$ characterized total information,
Shannon-based counterparts are well suited to enquire: What observable is this information about?

It turns out that the environment {\it as a whole} ``knows'' many 
observables of $\cS$, as is seen in Fig. 3a. By contrast, in Fig. 3b symmetry breaking is evident: The ridge of redundancy appears abruptly only when test observable $\sigma(\mu)$ 
and the preferred pointer observable $\sigma_z$ (that remains unperturbed by the environment)
nearly coincide. 

Why are pointer states favored? Commonsense  
says that, to be reproduced, state must survive copying.
This leads to a theorem~\cite{42,43} that only pointer states can be discovered from fractions of $\cE$. Other observables (such as $\sigma(\mu)$ in Fig.~3) can be deduced
only to the extent they are correlated with the pointer observable. So, fragments of the environment 
offer a very narrow, projective point of view. Redundant imprinting of some observables happens at the expense of their complements.

Structure of branching state betrays its origin and foreshadows ``collapse''. Starting from 
$\ket {\psi_{\cS}} = \sum_k^n\psi_k \ket {s_k}$,
$$\ket {\Psi_{\cS\cE}} = \sum_k^n\psi_k \ket {s_k} \ket {e^{(1)}_k}  
\dots \ket {e^{(\cN)}_k}= \sum_k^n \psi_k \ket {s_k} \ket {\varepsilon_k} \ \ (6)$$
branches grow to include $\cal N$ subsystems of $\cE$. Branch fragments
can be nearly orthogonal; $\Pi_{j=1}^{J} \bk {e_k^{(j)} }{ e_{k'}^{(j)}} \simeq \delta_{kk'}$ 
for large enough $J$. This means that a pointer state $ \ket {s_k}$ of $\cS$ can be determined (along with the rest of the 
branch) from a sufficiently long fragment (which may still be short compared to the length of 
the branch, $J\ll\cal N$).

In the huge Hilbert space $\cH_{\cS\cE}$ branching state
is a very atypical minimally entangled superposition of only $n$ product ``branches'' labelled by the
pointer states of the system. This is tiny compared to the dimension of $\cH_{\cS\cE}$ that exceeds $n$ by a factor exponential in $\cN$. 
This is why the two plots in Fig. 2 are so different: 
Branching state is, to a good approximation, a multi-system 
Schmidt decomposition, with long branch fragments constituting ``systems''. 
In a Schmidt decomposition, states of partners are in one-to-one correspondence. 
Thus, in Eq. (6), $\ket {s_k}$ implies $\ket {\varepsilon_k}$ (and, {\it vice versa}), and measuring a branch fragment $\cF$ can reveal the whole branch. 

Initial part of $I(\cS:\cF_f)$, Fig. 2, represent buildup of this correlation: When $f=0$, observer 
is ignorant of what branch he will find out, but the structure of the correlations within $\ket {\Psi_{\cS\cE}}$ 
leaves no doubt of what these branches are. Using Born's rule one could assign to them probabilities 
$p_k=|\psi_k|^2$ and the corresponding entropy $H_{\cS}$. Next section shows how one can deduce these probabilities 
without axiom (v) -- how symmetries of entanglement imply Born's rule. 

When observer measures enough of $\cE$, he finds out the branch (and what the state of $\cS$ is). 
Additional data are redundant. They only confirm 
what is already known. Probabilities associated with $\ket {\Psi_{\cS\cE}}$ are replaced with certainty
of a branch.
This transition from uncertainty (initial presence of many branches -- potential for
multiple outcomes) to certainty (once a sufficiently long branch fragment becomes known) accounts
for perception of ``collapse''. The initial, steeply rising, part of $I(\cS:\cF_f)$
``resolves'' it: Collapse is brief compared to the ensuing period 
of certainty about the outcome, as $f_{\delta} \ll 1$, but, nevertheless, not instantaneous.

Assumptions that lead from copying to preferred states can be relaxed. Thus, $\cE$ need not be initially pure~\cite{79}. Moreover, 
it suffices that the {\it records} (e.g., in the apparatus ${\cal A}$) are ``repeatably accessible''. 
Transfer of responsibility for repeatability from a quantum $\cS$ to a (still quantum) ${\cal A}$ allows one to model non-orthogonal measurement outcomes (POVM's): ${\cal A}$ entangles with the system, and then acts as ancilla. 
Its orthogonal pointer states $\ket{A_k}$ correlate with non-orthogonal $\ket{\varsigma_k}$ of ${\cS}$, $\sum_k\tilde\psi_k\ket{\varsigma_k}\ket{A_k}$. Interaction of $\cal A$ with the environment results in multiple copies of $\ket{A_k}$. The usual projective measurement implementation of POVM's (see e.g.~\cite{NC}) is now straightforward. Branches are labelled by $\ket{A_k}$. Indeed, we usually experience ``quantum jumps'' via an apparatus pointer.

Selection of the set of outcomes by the proliferation of information essential for Quantum Darwinism parallels Bohr's insistence~\cite{11} that a ``classical apparatus'' should determine the outcomes. However, 
it follows from purely quantum Eq. (5), and is caused by a unitary evolution responsible for the information transfer.  Nevertheless, as classical apparatus would, preferred pointer states designate possible future 
outcomes, precluding measurements of complementary observables 
or determining preexisting state of the system. 
Thus, information acquisition -- a copying 
process -- results in preferred states. 

Consensus between records deposited in fragments of $\cE$ looks like ``collapse''.  In this sense we have accounted for postulate (iv) using only very quantum postulates (i)-(iii). In particular, in deriving and analyzing Eq. (5) we have not employed Born's rule, axiom (v). 
We shall be therefore able to use our results as a starting point for such a derivation in the next section.

There was nothing nonunitary above --  unitarity was the crux of our argument, and orthogonality of branch seeds our main result. 
Relative states of Everett~\cite{25,26,22} come to mind. One could speculate about reality of branches with other outcomes. We abstain from this -- our discussion is interpretation-free, and this is a virtue. 
Indeed, ``reality'' or ``existence'' of universal state vector seems problematic. Quantum states acquire objective existence when reproduced in many copies. Individual states -- one might say with Bohr -- are mostly information, too fragile for objective existence.
And there is only one copy of the Universe. Treating its state as if 
it really existed~\cite{25,26,22} seems unwarranted and  ``classical''. 

\section{Probabilities from Entanglement}

\begin{figure*}[tb]
\begin{tabular}{l}
\vspace{-0.8in} 
\hspace{-0.3in}
\includegraphics[width=5.0in]{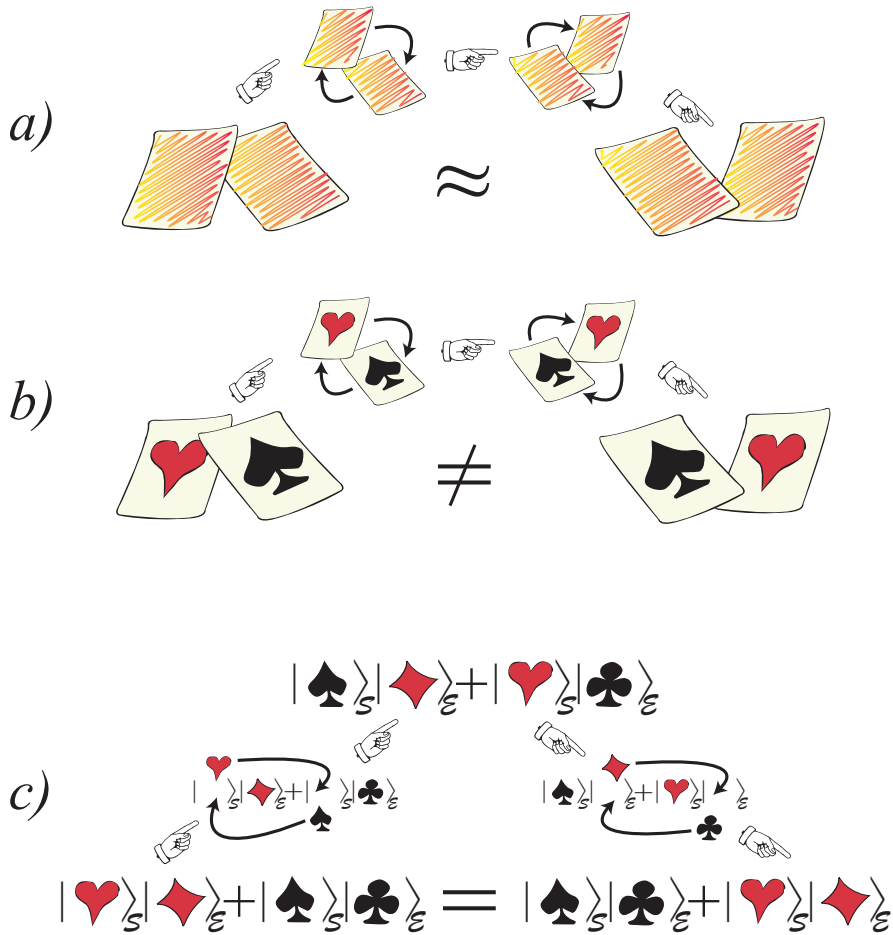}\\
\end{tabular}
\caption{{\it Probabilities and symmetry}: 
{\bf (a)} 
Laplace used
subjective ignorance 
to define probability. 
Player 
who does not know face values of the cards, but knows that one 
of  them is a spade will infer probability $p_\spadesuit ={ \frac 1 2}$ for the top card. 
{\bf (b)} The real physical state of the system is however altered by the swap, 
illustrating subjective nature of Laplace's approach, and demonstrating its unsuitability for physics.
{\bf (c)} {\it Perfectly known entangled states} have objective symmetries that allow one to rigorously deduce probabilities.
When two systems 
are maximally entangled 
as above, probabilities of Schmidt partners are equal, $p_{\heartsuit} = p_{\diamondsuit}$, and $p_{\spadesuit}=p_{\clubsuit}$.  After a swap $u_{\cS}=\ket \spadesuit \bra \heartsuit + \ket \heartsuit \bra \spadesuit$ in $\cS$, the resulting state $\ket \spadesuit \ket \diamondsuit + \ket \heartsuit \ket \clubsuit $ must have $p'_{\spadesuit} = p_{\diamondsuit}$, and $p'_{\heartsuit}=p_{\clubsuit}$. (We `primed' 
probabilities in $\cS$,
as it was acted upon by a swap, so they might have changed.) A counterswap
$u_{\cE}=\kb {\diamondsuit} {\clubsuit} + \kb  {\clubsuit} {\diamondsuit} $
in $\cE$ restores the original entangled state, proving that $p'_{\heartsuit }=p_{\heartsuit }$ 
and $p'_{\spadesuit} =p_{\spadesuit}$, after all (as counterswap $u_{\cE}$ leaves $\cS$ untouched). This sequence 
of equalities implies $p_\spadesuit = p_{\diamondsuit}= p_\heartsuit$, so that $p_\spadesuit =  p_\heartsuit=\frac 1 2$, as probabilities in $\cS$ must add up to 1.}
\label{cards}
\end{figure*} 

Observer prepared $\cS$ in a state $\ket {{\psi}_{\cS}}$, but wants to measure observable 
with eigenstates $\{\ket {s_k}\}$. This will lead to entangled $\ket {\Psi_{\cS\cE}}$ with branch structure, Eq. (6).
Pointer states $\{\ket {s_k}\}$ define the outcomes, but, as yet, observer 
has not measured $\cE$, and does not know the result. Given $\ket {\Psi_{\cS\cE}}$, what is the probability of, say, $\ket {s_{17}}$?

To derive it we cannot use reduced density matrices, Eqs.~(1,2). Tracing out is 
averaging~\cite{Landau,59,NC} -- it relies on $p_k=|\psi_k|^2$, Born's rule we want 
to derive. We have imposed that ban while deriving and analyzing Eq. (5), but relaxed it to plot Fig. 3.
Now we reimpose it again. So, Born's rule and standard tools of decoherence are off limits -- using them courts 
circularity. Our derivation will rest instead on {\it certainty} and {\it symmetry}, cornerstones that mark two 
extremal cases of probability. 

The case of certainty was just settled without Born's rule using Eq. (5). When one 
re-measures an observable, the same outcome will be seen again. Thus, when $\{\ket {s_k}\}$ includes 
$\ket {{\psi}_{\cS}}$ (e.g., $\ket {{\psi}_{\cS}}=\ket {s_{17}}$), newly added copies just extend the branch
already correlated with observer's state,
and the outcome is certain; 
$p_{17}=1$. Certainty of correlations between partners in 
Schmidt decomposition, Eq. (6) is another important example.

Certainty seems trivial but is important. Confirmation that a state ``is what it is'' -- postulate (iii) -- is a part of standard quantum lore~\cite{23}. We re-affirmed it, but with a key insight: Redundancy allows observers to discover (and not just confirm) that $\cS$ is in a certain pointer state.
 
We now turn to the opposite case 
of complete indeterminacy.
Its connection with symmetry was noted by Laplace. He wrote: {\it ``The theory of chance consists in reducing all the events ... 
to a certain number of cases that are equally possible...
The ratio of this number to that of all the cases possible is the measure of probabilityÉ''}~\cite{40}. 
Figure 4 illustrates how this classical intuition yields -- far more convincingly --- quantum probabilities.

Symmetry is probed by invariance. Transformations that respect it take system between states that 
exhibit no measurable differences. For example, change of phase in the coefficients in the 
Schmidt decomposition $\ket {\Psi_{\cS\cE}} = \sum_k^n \psi_k \ket {s_k} \ket {\varepsilon_k} $ cannot influence the state of $\cS$: 
It is induced by $u_{\cS}=e^{i\phi_k} \kb {s_k}{s_k}$, local unitary on $\cS$, that
can be ``undone'' by 
$u_{\cE}=e^{-i\phi_k} \kb {\varepsilon_k}{\varepsilon_k}$ on $\cE$, or; 
$$u_{\cS}\otimes {\bf 1}_{\cE}\ket {\Psi_{\cS\cE}} = \ket {\Phi_{\cS\cE}}; \ {\bf 1}_{\cS}\otimes u_{\cE}\ket {\Phi_{\cS\cE}}=\ket {\Psi_{\cS\cE}} \eqno(7) $$
So, 
phases of $\psi_k$ cannot matter for a
local state or influence probabilities in $\cS$. This symmetry, Eq. (7), is the {\it en}tanglement-assisted 
in{\it variance} or {\it envariance}~\cite{76,78}. 

Such loss of phase significance for $\cS$ entangled with $\cE$ implies decoherence~\cite{78}. 
We arrived at its essence using envariance, without reduced density matrices, trace, etc.

We now use phase envariance to show that equal absolute values of the coefficients $\psi_k$ imply  equal probabilities. For equal $|\psi_k|$ any
orthogonal basis of $\cS$ is ``Schmidt'' (i.e., has an orthogonal partner in $\cE$). Thus, $\ket {\bar \varphi_{\cS\cE}} =\frac {{\ket 0}_{\cS}{\ket 0}_{\cE}+{\ket 1}_{\cS}{\ket 1}_{\cE}} {\sqrt 2} =\frac {{\ket +}_{\cS}{\ket +}_{\cE} + {\ket -}_{\cS}{\ket -}_{\cE}} {\sqrt 2} $, where $\ket \pm = \frac {\ket 0 \pm \ket 1} {\sqrt 2}$.
Sign change induced by $e^{i \pi} \kb {-}{-} $ acting on $\cS$
produces $\ket {\bar \eta_{\cS\cE}}=\frac {{\ket +}_{\cS}{\ket +}_{\cE} - {\ket -}_{\cS}{\ket -}_{\cE}} {\sqrt 2} =\frac {{\ket 1}_{\cS}{\ket 0}_{\cE}+{\ket 0}_{\cS}{\ket 1}_{\cE}} {\sqrt 2} $. In other words, 
one can swap ${\ket 0}_{\cS}$ with ${\ket 1}_{\cS}$ by rotating phase in a $\ket \pm$ basis by $\pi$. Yet, we just saw that phases of Schmidt coefficients do not matter for the state
of $\cS$, so probabilities of 0 and 1 in $\cS$ must have remained the same. Moreover, probabilities of paired up Schmidt states are equal, so that $p_{\cS}(0)=p_{\cE}(0)$ in $\ket {\bar \varphi_{\cS\cE}}$ 
and $p_{\cS}(1)=p_{\cE}(0)$ in $\ket {\bar \eta_{\cS\cE}}$. 
Hence, $p_{\cS}(0)=p_{\cS}(1)=\frac 1 2$, where we assumed that probabilities add up to 1.
 
In contrast to Laplace's subjective ``ignorance-based'' approach, we obtained {\it objective probabilities for a completely known entangled state}. Phase envariance implied equiprobability in $\cS$. 
To paraphrase {\it Beatles}, ``All you need is phase...''. We rotated phases of the coefficients to induce a swap in a complementary basis. Another proof (that implements swap more directly) is given in Fig. 4. 

This equiprobability case  is the difficult part of the proof. Instead of subjectivity 
(that undermined applicability of Laplace's approach to physics) we relied on objective symmetries 
of entangled quantum states. This was made possible by  the nature of quantum states of composite systems. Classically, pure states have structure of a Cartesian product -- knowing the whole implies knowledge of each subsystem. In quantum theory they are tensor products -- 
one can know state of the whole, and thus know nothing about parts, as envariance shows. 

This was the basis of our proof of equiprobability. We assumed unitarity. Moreover, we assumed; 
(1) {\it When a system is not acted upon by a unitary transformation, its state remains unaffected. }
This state is a property of $\cS$ alone, so; (2) {\it Predictions regarding measurement outcomes 
on $\cS$ (including their probabilities) can be inferred from the state of $\cS$}. Last not least;
(3) {\it When $\cS$ is entangled with other systems (e.g., the environment) the state of $\cS$ alone is determined by the state of the whole $\cS\cE$}. 

These ``facts of life'' are accepted properties 
of systems and states, but given the fundamental nature of our discussion it seems a good idea 
to make them explicit~\cite{78}.

For instance, to establish independence from phases of the coefficients $\psi_k$ we noted that the state of $\cS$ is unaffected by the unitaries $u_{\cS}$ diagonal in Schmidt basis
acting on $\cS$ (like changes of Schmidt coefficient phases) that would normally affect isolated $\cS$:
The global state $\Psi_{\cS\cE}$ is restored by $u_{\cE}$. Thus, by fact (3), so is local state of 
$\cS$. However, this is done by a unitary ``countertransformation'' acting solely 
on $\cE$.
Hence, by fact (1), state of $\cS$ must have been unaffected by $u_{\cS}$ in the first place. So, by fact (2), phases of  $\psi_k$ cannot change
outcomes of any measurement on $\cS$. Equiprobability follows.

One can now derive Born's rule, $p_k=|\psi_k|^2$, with straightforward algebra from the above two simple cases of complete certainty ($p_k=1$) and equiprobability ($p_k= \frac 1 n$): 
The general case can be always reduced to the case  case of equal coefficients by ``finegraining'' (see {\bf Box}). 

The origin of probability is a fascinating problem that is older than quantum measurement problem, 
and is forgotten primarily because it is so old. We have seen how quantum physics sheds a new,
very fundamental, light on probability. We cannot do justice to the history of this subject here,
but Ref.~\cite{Aul} provides a basic overview and exhaustive set of references. In particular, envariant derivation is very different from the classic proof of Gleason~\cite{Gle} in that it sheds light on the physical significance of the resulting measure. Moreover, it does not assume probabilities are additive (except to posit that probability of an event and its complement are certain, i.e., to establish normalization; see {\bf Box} and Ref.~\cite{78,Z07}). Bypassing additivity of probabilities is essential when dealing with a theory with another principle of additivity -- the quantum superposition principle -- which trumps additivity of probabilities or at least classical intuitiions about it (e.g., in the double-slit experiment). Discussion of the 
implications of envariance has already started, with~\cite{SF, Bar}, and~\cite{52} providing 
insightful commentary.

\medskip
\noindent{\bf BOX}

We show here how ``finegraining'' reduces the case of arbitrary $\psi_k$ to equiprobability.
To illustrate general strategy consider state 
in a
2D Hilbert space ${\cal H}_{\cS}$ of $\cS$ spanned by orthonormal
$\{|0\rangle,|2\rangle\}$ and (at least) 3D  ${\cal H}_{\cE}$:

\noindent$ \ \ \ \ \ \  |\psi_{\cal SE} \rangle \ \propto \ \sqrt{\frac 2 3}~|0\rangle_{\cS}|+ \rangle_{\cE} \ \ + \ \  \sqrt{\frac 1 3} ~|2\rangle_{\cS}|2\rangle_{\cE}  \ .  $

\noindent The state $|+\rangle_{\cE}=\frac {|0\rangle_{\cE}+|1\rangle_{\cE}} {\sqrt 2}$ exists
in (at least 2D) subspace of ${\cal E}$ orthogonal to $|2\rangle_{\cE}$, i.e., 
$\langle0|1\rangle=\langle0|2\rangle=\langle1|2\rangle=\langle+|2\rangle=0$.
We know we can ignore phases.

To reduce $|\psi_{\cal SE}\rangle$
to equal coefficients case
we ``extend it'' to a state $|\bar \Psi_{\cal SEC}\rangle$ by letting $\cE$ act on
an ancilla ${\cal C}$. ($\cS$ is not acted upon, so, by fact (1), probabilities
for $\cS$ cannot change.) This can be done
by a generalization of controlled-not 
acting between ${\cal E}$ (control) and ${\cal C}$ (target), so that
(in obvious notation) $|k\rangle|0'\rangle \Rightarrow |k\rangle|k'\rangle$, leading to
\smallskip

\noindent$ 
\sqrt 2 \ket 0 \ket + \ket {0'} + \ket 2 \ket 2 \ket {0'}
\Rightarrow \sqrt 2 |0\rangle{{|0\rangle|0'\rangle + |1\rangle|1'\rangle} \over \sqrt 2}  + |2\rangle|2\rangle|2'\rangle .
$
\smallskip

\noindent Above, and from now on we skip subscripts: The state of ${\cal S}$ will be listed first,
and the state of ${\cal C}$ will be primed. 

The cancellation of $\sqrt 2$ yields an equal coefficient state:
$$ |\bar \Psi_{\cal SCE}\rangle \propto |0,0'\rangle|0\rangle + |0,1'\rangle|1\rangle 
 + |2,2'\rangle|2\rangle  \ . $$

\noindent We have combined $\cS$ and ${\cal C}$ in a single ket and (below) we shall swap states of  ${\cal SC}$ as if it was a single system.

Clearly, 
this is a Schmidt decomposition of ($\cS \cal C) {\cal E}$. Three orthonormal product states have
coefficients with the same absolute value. 
Therefore, they can be envariantly swapped. 
Thus, the probabilities of 
states $|0\rangle|0'\rangle, \ |0\rangle|1'\rangle,$ 
and $|2\rangle|2'\rangle$ are all equal. By normalization they are $\frac 1 3$. So, 
probability of detecting state $|2\rangle$ of $\cS$ is $ \frac 1 3$. Moreover, $\ket 0$ and $\ket 2$ 
are the only two outcome states for $\cS$. It follows that probability of 
$|0\rangle$ must be $\frac 2 3 $;

\noindent $ \ \ \ \ \ \ \ \ \ \ \ \ \ \ \ \ \ \ \  p_0 =  \frac 2 3; \ \ p_2= \frac 1 3 \ .  $

\noindent This is Born's rule. We have just seen why the amplitudes in the initial $|\psi_{\cal SE}\rangle$ ``get squared'' to yield probabilities.

Note that we have avoided assuming additivity of probabilities: $ p_0 = \frac 2 3$ not because 
it is a sum of two fine-grained alternatives for $\cS\cE$, each with probability of $ \frac 1 3$, but rather because there are only two (mutually exclusive and exhaustive) alternatives for $\cS$; $\ket 0$ and 
$\ket 2$, and $p_2= \frac 1 3$. Therefore, by normalization, $ p_0 = 1-  \frac 1 3$.
Probabilities of Schmidt states can be added because of the loss of phase coherence that follows directly from phase envariance established earlier (see also Ref.~\cite{76,78}).

Extension of this proof to the case where probabilities are commensurate is conceptually straightforward
but notationally cumbersome. The case of non-commensurate probabilities is settled with an appeal to continuity. Frequency of the outcomes can be also deduced, allowing one to
establish connection with the familiar relative frequency approach to probabilities~\cite{76,78,Z07}, but in a quantum setting probability arises as a consequence of symmetries of a single entangled state.

We end by noting that the finegraining discussed above does not need to be carried out experimentally each time probabilities are discussed: Rather, it is a way to deduce a measure that is consistent with the geometry of the Hilbert spaces using entanglement as a tool. Still, given fundamental implications of envariance experimental tests would be most useful.

\section{Discussion}

We derived the two controversial quantum postulates from the first three. We have 
thus seen how classical domain of the Universe arises from the superposition principle (postulate (i)) and unitarity (postulate (ii)) as well as rudimentary assumptions about information flows
(postulate (iii)), and a few basic facts about states of composite quantum systems (including their tensor nature, often cited as additional ``axiom (0)''). 

The essence of the measurement problem -- accounting for axioms (iv) and (v) -- has been largely settled. It is of course likely one may be able to clarify assumptions and simplify proofs. Much work remains to be done on Quantum Darwinism and envariance.  Nevertheless, nature
of the quantum-classical correspondence has been clarified.



Physicists take it for granted that even hard problems are solved by a single good idea. Therefore, when 
a {\it single} idea does not do the {\it whole} job, often our first instinct is to dismiss it. Measurement problem 
does not fall into this ``single idea'' category. Several ideas, applied in the right order, led to advances described here. Logically, we may well have started with the derivation of Eq. (5) and the analysis of 
quantum jumps. Their randomness leads to probabilities. And symmetries of entangled states (that 
arise in decoherence and Quantum Darwinism) allow one to derive Born's rule. As we have seen, 
phase envariance is (nearly) ``all you need''. With probabilities at hand one has then every right to 
use reduced density matrices to analyze Quantum Darwinism and decoherence. 


Our presentation was ``historical''. We started with decoherence, and used it to introduce Quantum 
Darwinism. Analysis of copying essential to information flows in both of these phenomena led to  quantum jumps. This in turn motivated entangelment-based derivation of Born's rule. 
Quantum Darwinism -- upgrade of $\cE$ to a communication channel from a mundane role it played 
in decoherence -- tied together all of the other developments. This order had the advantage of making 
motivations clear, but it is different from more logical presentation where postulates (i)-(iii) are the 
starting point (strategy followed in \cite{Z07}).

The collection of ideas discussed here allows one to understand how ``the classical'' emerges 
from the quantum substrate staring from more basic assumptions than decoherence. We have 
bypassed a related question of why is our Universe quantum to the core. The nature of quantum state
vectors is a part of this larger mystery. Our focus was not on what quantum states {\it are}, but on 
what they {\it do}.  Our results encourage a view one might describe (with apologies to Bohr) as ``complementary''. Thus, $\ket \psi$ is in part information (as, indeed, Bohr thought), but also the obvious quantum object to explain ``existence''. We have seen how Quantum Darwinism accounts for the transition from quantum fragility (of information) to the effectively classical 
robustness. One can think of this transition as ``It from bit'' of John Wheeler~\cite{JAW}. 

In the end one might ask: ``How Darwinian is Quantum Darwinism?''. Clearly, there is survival of the fittest, and fitness is defined as in natural selection -- through the ability to procreate. The no-cloning theorem implies competition for resources -- space in $\cE$ -- so that only pointer states can multiply (at the expense of their complementary competition). There is also another aspect of this competition:  Huge memory available in the Universe as a whole is nevertheless limited. So the question arises: What systems get to be ``of interest'', and imprint their state on their obliging environments, and what are the environments? Moreover, as the Universe has a finite memory, old events will be eventually 
``overwritten'' by new ones, so that some of the past will gradually cease to be reflected in the present record. And if there is no record of an event, has it really happened? 
These questions seem 
far more interesting than deciding closeness of the analogy with natural selection~\cite{Darwin}. They suggest one more question: Is Quantum Darwinism (a process of multiplication of information about certain favored states that seems to be a ``fact of {\it quantum} life'') in some way behind the familiar natural selection? I cannot answer this question, but neither can I resist raising it.

\noindent Acknowledgments: I am grateful to Robin Blume-Kohout, Fernando Cucchietti, Juan Pablo Paz, David Poulin, Hai-Tao Quan, Michael Zwolak for stimulating discussions. This research was supported by an LDRD grant at Los Alamos and, in part, by FQXi. 

\end{document}